\documentclass[preprint,showpacs,amsmath,amssymb,pra,floatfix,superscriptaddress]{revtex4-1}
\usepackage{graphicx}
\usepackage{dcolumn}
\usepackage{bm}
\usepackage{xspace}
\usepackage{hyperref}
\usepackage{color}

\def\rucl{$\alpha$-RuCl$_3$\xspace}
\providecommand*{\et}{\emph{et\,al.}\xspace}%
\def\yzgo{YbZnGaO$_4$\xspace}
\def\ymgo{YbMgGaO$_4$\xspace}
\begin{document}

\title{Recent progress on magnetic-field studies on quantum-spin-liquid candidates}
\author{Zhen~Ma}
\author{Kejing~Ran}
\author{Jinghui~Wang}
\author{Song~Bao}
\author{Zhengwei~Cai}
\author{Shichao~Li}
\affiliation{National Laboratory of Solid State Microstructures and Department of Physics, Nanjing University, Nanjing 210093, China}
\author{Jinsheng~Wen}
\email{jwen@nju.edu.cn}
\affiliation{National Laboratory of Solid State Microstructures and Department of Physics, Nanjing University, Nanjing 210093, China}
\affiliation{Collaborative Innovation Center of Advanced Microstructures, Nanjing University, Nanjing 210093, China}


\begin{abstract}
Quantum spin liquids (QSLs) represent a novel state of matter in which quantum fluctuations prevent conventional magnetic order from being established, and the spins remain disordered even at zero temperature. There have been many theoretical developments proposing various QSL states. On the other hand, experimental movement was relatively slow largely due to limitations on the candidate materials and difficulties in the measurements. In recent years, the experimental progress has been accelerated. In this topical review, we give a brief summary of experiments on the QSL candidates under magnetic fields. We arrange our discussions by two categories: i) Geometrically-frustrated systems, including triangular-lattice compounds YbMgGaO$_4$ and YbZnGaO$_4$, $\kappa$-(BEDT-TTF)$_2$Cu$_2$(CN)$_3$, and EtMe$_3$Sb[Pd(dmit)$_2$]$_2$, and kagom\'{e} system ZnCu$_3$(OH)$_6$Cl$_2$; ii) the Kitaev material $\alpha$-RuCl$_3$. Among these, we will pay special attention to $\alpha$-RuCl$_3$, which has been intensively studied by our and other groups recently. We will present evidence that both supports and unsupports the QSL ground state for these materials, based on which we give several perspectives to stimulate further research activities.
\end{abstract}
\pacs{61.05.fg, 75.10.Kt, 75.50.Lk}

\maketitle
\section{Introduction}

\begin{figure}[htb]
\centering
\includegraphics[width=0.98\linewidth,trim=0mm 0mm 0mm 0mm]{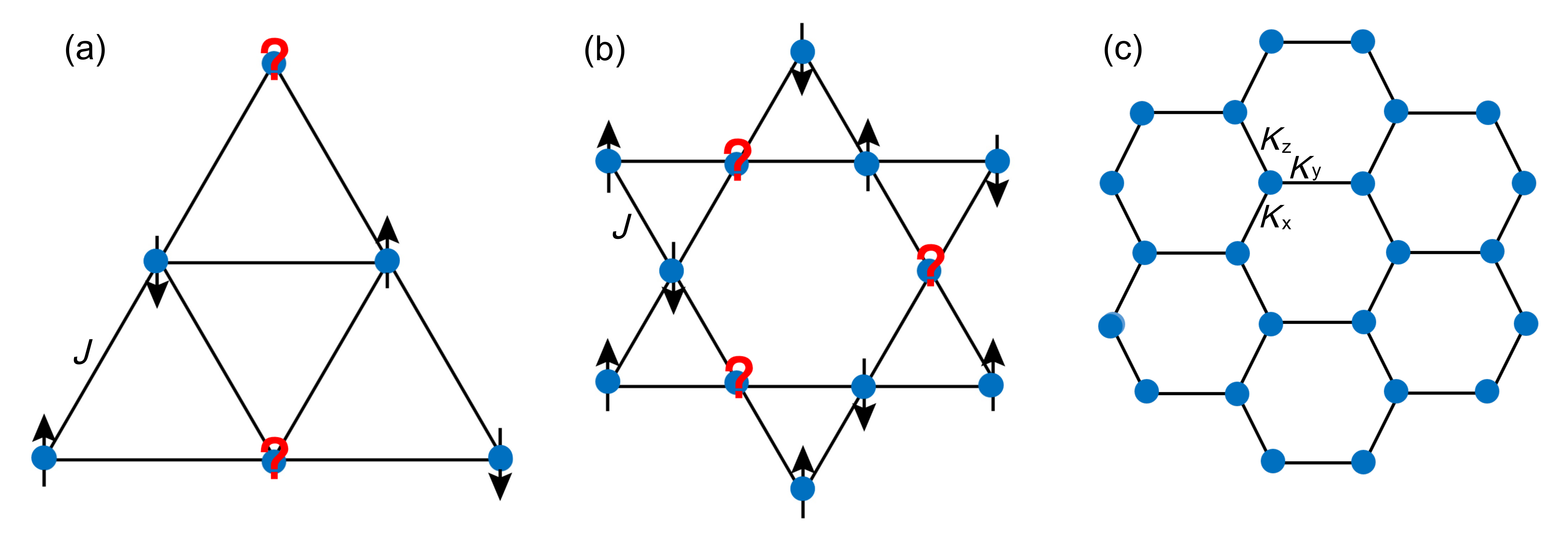}\\[5pt]
\caption{\label{qslstructures}{Schematics of two-dimensional (a) triangular, (b) kagom\'{e}, and (c) honeycomb structures where quantum spin liquids can be realized. Arrows and question marks represent spins and geometrical frustration, respectively. $J$ is the Heisenberg exchange interaction, and $K_{x,y,z}$ are Kitaev interactions along three bonds.}}
\end{figure}

Generally, a system tends to lower its symmetry to be in the low-energy state. As a consequence, a magnetic material will break certain symmetry according to Landau's theorem, and the magnetic moments carrying by the electron spins will form an ordered pattern at low temperatures\cite{Magnetism,qtMagnetism}. Excitations associated with these ordered spins are conventional magnons with spin $S=1$~(refs~\cite{Neel1985,PhysRev.115.2,PhysRevLett.99.127004,PhysRevLett.82.3899,PhysRevLett.68.1766,PhysRevB.50.10048,PhysRevLett.69.2590}).
However, in systems with small spin and strong quantum fluctuations, such a conventional order can be avoided, leading to a quantum-spin-``liquid"~(QSL) state\cite{qslfirst}. Now, it is known that geometrical frustration, a situation where the antiferromagnetic Heisenberg exchange interactions cannot be satisfied simultaneously among different sites in triangular~[Fig.~\ref{qslstructures}(a)] or kagom\'{e} lattice~[Fig.~\ref{qslstructures}(b)], can result in strong quantum fluctuations\cite{nature464_199,Lee1306,PhysRevB.65.165113}. In 1973, Anderson proposed the resonant-valence-bond (RVB) model on the triangular lattice for the QSL state. It is a superposition of all possible configurations of the singlets formed by any of the two strongly interacting spins\cite{Anderson1973153,Moessner01022002,PhysRevLett.61.365,doi:10.1080/14786439808206568}.
The elementary excitations in QSLs are fractionalized quasiparticles, $e.g.$, charge-free $S=1/2$ spinons, fundamentally different from conventional magnons\cite{Anderson1973153,Moessner01022002,PhysRevLett.61.365,doi:10.1080/14786439808206568}.

The QSL state defined by the RVB model does not have an exact solution. In 2006, Kitaev constructed an exactly-solvable $S=1/2$ model on the honeycomb lattice [Fig.~\ref{qslstructures}(c)]. This model, as in Eq.~\ref{kitaevmodel}, is named the Kitaev model.
\begin{equation}
\label{kitaevmodel}
H=-K_x\sum_{x\text{-bonds}}S_i^xS_j^x-K_y\sum_{y\text{-bonds}}S_i^yS_j^y-K_z\sum_{z\text{-bonds}}S_i^zS_j^z
\end{equation}
Here, $H$ is the Hamiltonian, and $K_{x,y,z}$ are the nearest-neighbor Kitaev interactions of the $x$, $y$, and $z$ bonds on the honeycomb lattice.
QSLs defined by this model is termed the Kitaev QSL. Unlike QSLs arising from geometrical frustration, the Kitaev QSL results from bond-dependent interactions that frustrate spin configurations on a single site\cite{Kitaev2006Anyons}. Initially, Kitaev model was treated as a toy model since the anisotropic spin interactions are unrealistic in a spin-only system. Later, it was suggested that in Mott insulators with strong spin-orbital coupling~(SOC), the anisotropic Kitaev interaction may be achievable due to the spatial anisotropy of the orbitals\cite{prl102_017205}. As such, our discussions on the QSL candidates will be based on two categories: geometrical-frustration-induced ones and Kitaev QSL candidates.

Besides the rich and exotic physics of QSLs, they also hold promising application potentials, for example, quantum computation via braiding the non-Abelian anyons in these materials\cite{Kitaev2006Anyons,Kitaev20032,0034-4885-80-1-016502,RevModPhys.80.1083,Barkeshli722}.
Furthermore, understanding QSLs may help understand the mechanism of high-temperature superconductivity\cite{mullerlbco,anderson1,PhysRevLett.58.2790,Baskaran1987973,PhysRevB.35.8865}. For these reasons, research on QSL has been surging in the past few decades. There have been many review articles summarizing the progress on QSLs already\cite{nature464_199,arcmp2_167,0034-4885-80-1-016502,arcmp5_57,RevModPhys.89.025003,0034-4885-78-5-052502,0034-4885-74-5-056501,doi:10.1080/14786430601080229,RevModPhys.88.041002,arcmp7_195,RevModPhys.82.53,arcmp3_35,1742-6596-320-1-012004,doi:10.1143/JPSJ.79.011001,imai2016quantum,0953-8984-29-49-493002}.
In this topical review, we will restrict our discussions to the measurements under magnetic fields only.

In general, an external magnetic field can be detrimental to the QSL phase, as the field may induce symmetry breaking. Ultimately, when the field is strong enough, all the spins will be polarized and the moments will align with the field direction---the system is then a ferromagnet. But on the other hand, in the fully polarized state, one can extract the exchange interactions from the spin-wave excitation spectra obtained by inelastic neutron scattering (INS) measurements and understand the magnetic ground state in zero field. Investigating the magnetic excitations under fields with techniques such as nuclear magnetic resonance (NMR), muon spin relaxation ($\mu$SR), electron spin resonance, and terahertz spectroscopy also provides key information on the interactions underlying the exotic states of the QSL candidates. Furthermore, studying the field evolution of the thermal transport properties can also provide insights into the QSL physics. Finally, in available systems where the Kitaev physics is relevant, there are other non-Kitaev terms setting in at low temperatures, resulting in an ordered phase instead of the Kitaev QSL. In some of these materials, such as \rucl, applying a magnetic field suppresses the non-Kitaev interactions and drives the system into a possible QSL state. In this article, we will summarize the results from these magnetic-field experiments on the i) Geometrically-frustrated systems, including triangular-lattice compounds YbMgGaO$_4$ and YbZnGaO$_4$, $\kappa$-(BEDT-TTF)$_2$Cu$_2$(CN)$_3$, and EtMe$_3$Sb[Pd(dmit)$_2$]$_2$, and kagom\'{e} system ZnCu$_3$(OH)$_6$Cl$_2$; ii) the Kitaev material $\alpha$-RuCl$_3$, which has been subject to intensive investigations by our and other groups recently. We first present experimental evidence for each of these materials, then make discussions based on these results, and raise several questions in the end.

\section{Geometrically-frustrated systems}

\subsection{\ymgo and \yzgo}

\begin{figure}[htb]
\centerline{\includegraphics[width=0.98\linewidth]{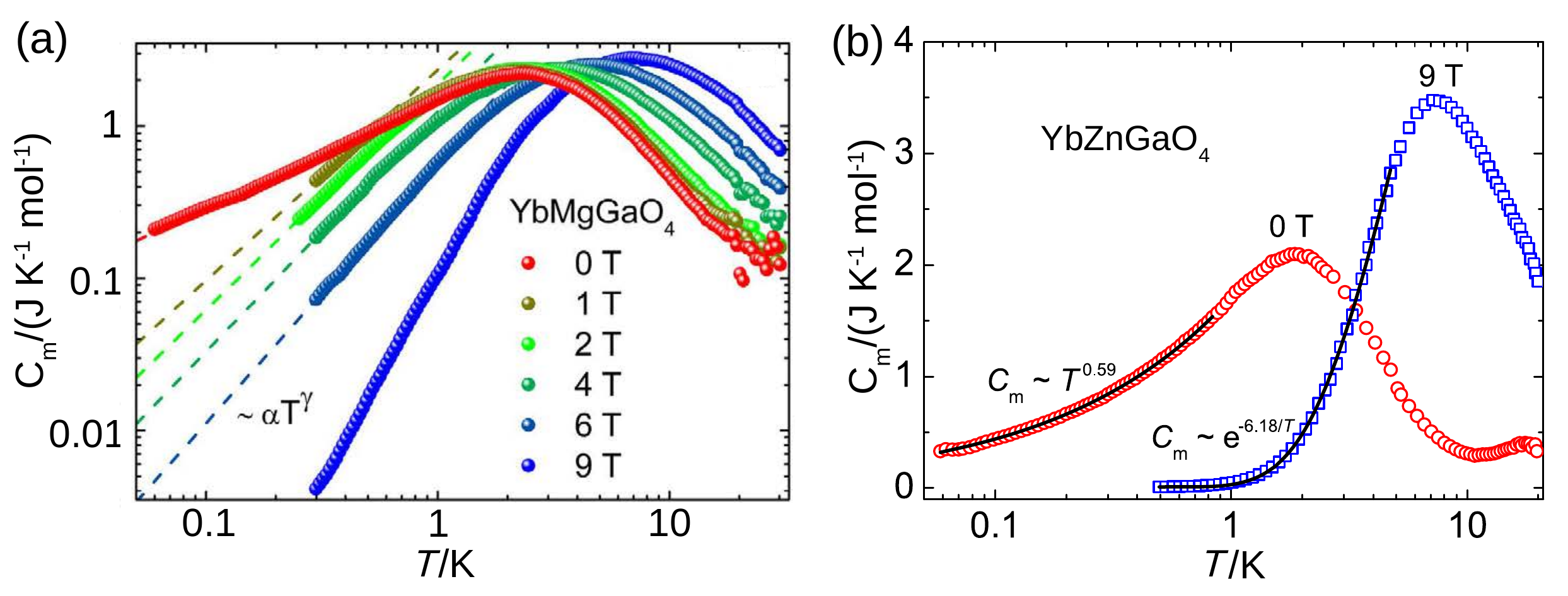}}
\caption{(a) Temperature dependence of the magnetic specific heat of \ymgo under different magnetic fields. Dashed lines indicate power-law fits to the low-temperature data. From ref.~\cite{sr5_16419}. (b) Magnetic specific heat of \yzgo measured under zero and 9-T fields. Solid lines are fits to the data. From ref.~\cite{Ma2017Quantum}.}
\label{Yb1}
\end{figure}

\ymgo with the quasi-two-dimensional triangular-lattice structure has been reported to be a promising QSL candidate recently\cite{sr5_16419,prl115_167203,np13_117,nature540_559,PhysRevLett.117.097201,PhysRevLett.118.107202,arXiv:1704.06468}.  It has a negative Curie-Weiss temperature of $\sim-4$~K\cite{sr5_16419,prl115_167203} but does not show a long-range magnetic order down to 30~mK, indicating strong geometrical frustration\cite{nature540_559,np13_117}. The magnetic specific heat under different fields for \ymgo is shown in Fig.~\ref{Yb1} (a)\cite{sr5_16419}. In zero field, there is a broad hump at 2.4~K, instead of a sharp $\lambda$-type peak expected for a well-defined phase transition. The magnetic specific heat is large at low temperatures, suggesting a gapless ground state. By applying magnetic fields up to 9~T, the system becomes a fully-polarized ferromagnetic state and opens a spin gap of $\sim$8.26~K\cite{PhysRevLett.117.267202,nature540_559,sr5_16419}. \yzgo, a compound isostructural to \ymgo, shows very similar behaviors. In Fig.~\ref{Yb1} (b), the magnetic specific heat results of \yzgo in zero and 9-T fields are shown. Similarly, the results indicate a gapless state in zero field, and  a gapped state with a gap size of $\sim$6.18~K in a 9-T field\cite{Ma2017Quantum}.

\begin{figure}[htb]
\centerline{\includegraphics[width=0.9\linewidth]{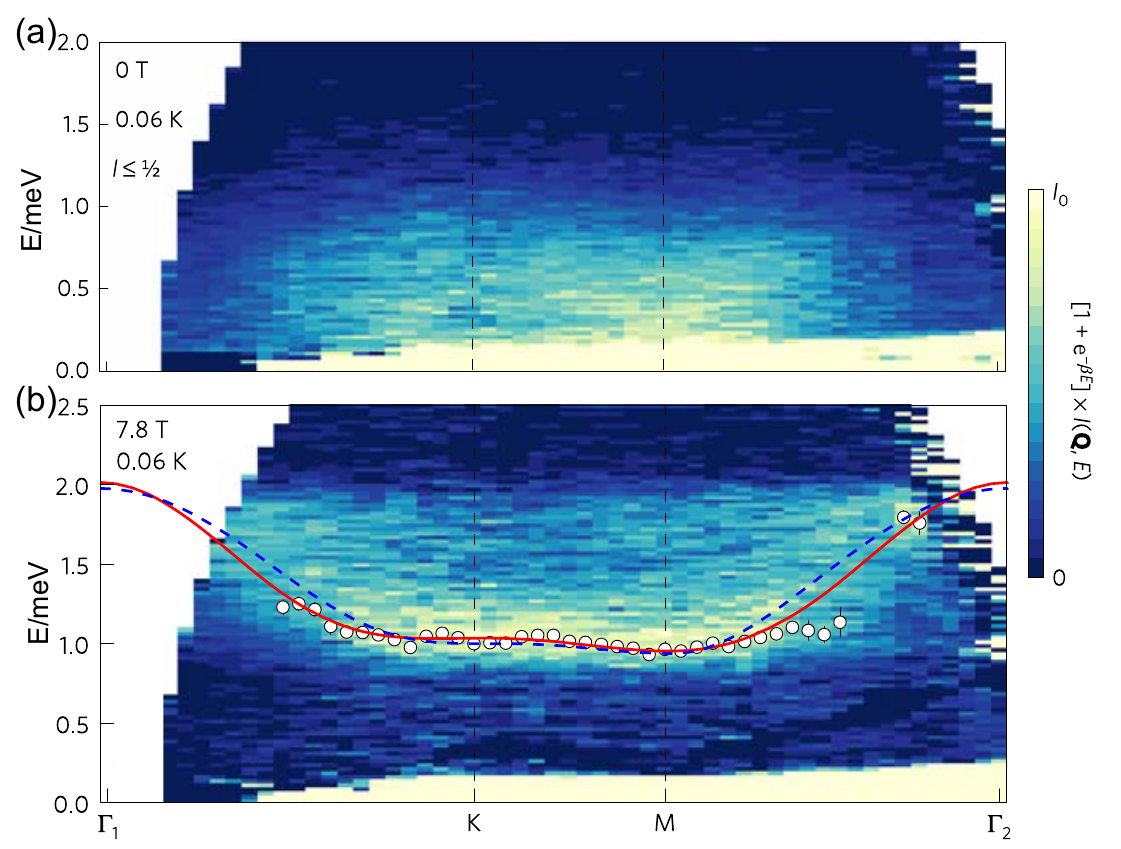}}
\caption{Magnetic excitation spectra along high-symmetry directions under (a) zero, and (b) 7.8-T fields applied parallel to the $c$ axis. Curves are fits to the spin-wave excitations using the anisotropic spin model. From ref.~\cite{np13_117}.}
\label{Yb2}
\end{figure}

\begin{figure}[htb]
\centerline{\includegraphics[width=0.65\linewidth]{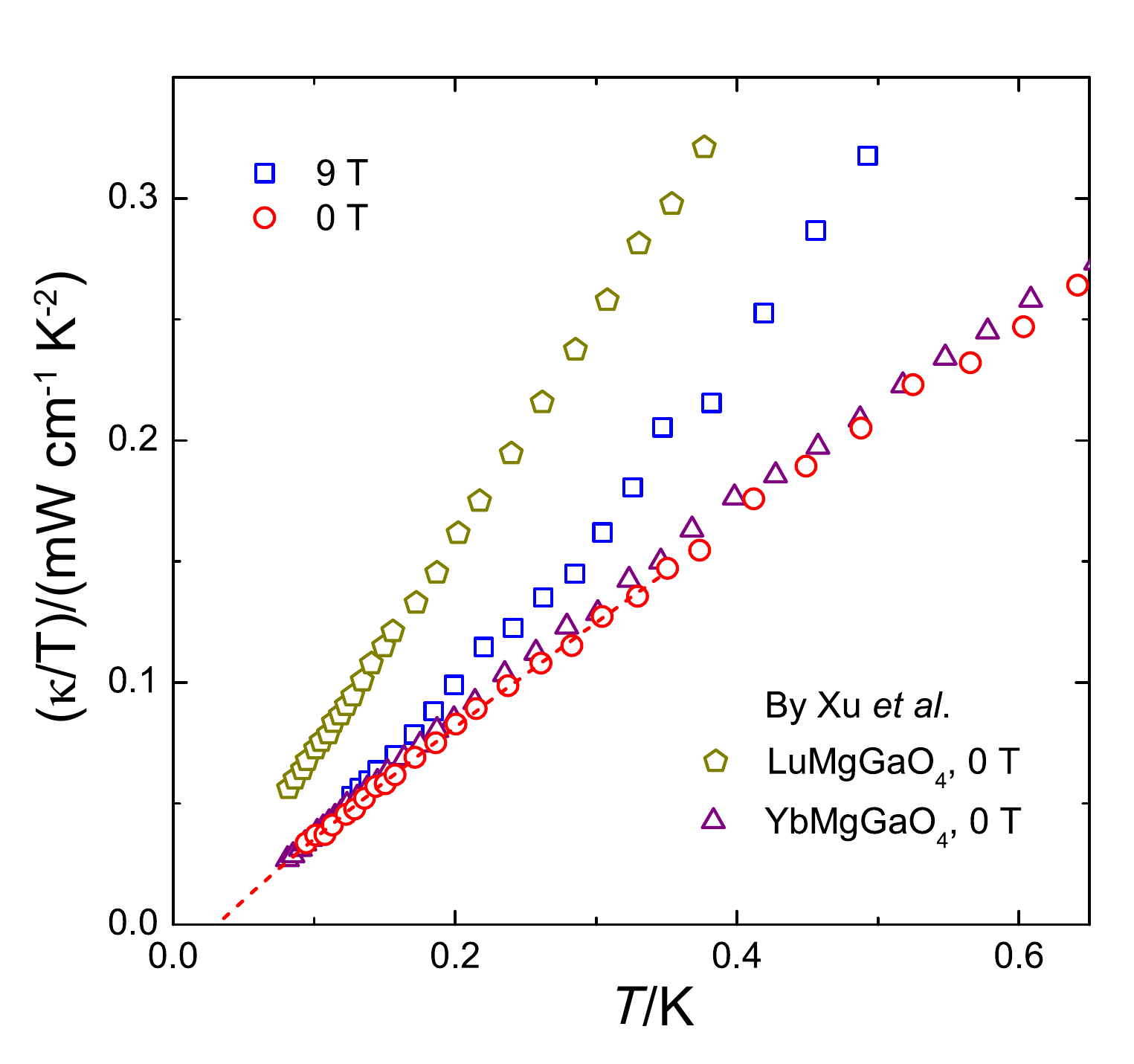}}
\caption{The in-plane thermal conductivity of a \yzgo single crystal under zero and 9-T fields applied parallel to the $c$ axis. For comparison, the thermal conductivity of \ymgo and LuMgGaO$_4$ reported in ref.~\cite{PhysRevLett.117.267202} is also plotted. From ref.~\cite{Ma2017Quantum}.}
\label{Yb3}
\end{figure}

In zero field, the magnetic excitation spectra for both \ymgo\cite{nature540_559,np13_117} and \yzgo\cite{Ma2017Quantum} exhibit as a broad ``continuum" in INS measurements, as shown in Fig.~\ref{Yb2}(a)\cite{np13_117}. Since QSLs are characterized by spin excitations carrying fractional quantum numbers, {\emph e.g.}, deconfined spinons, such observations of the continuum have been taken to be strong evidence for a QSL state in YbMgGaO$_4$\cite{nature540_559,np13_117}. Applying a magnetic field on the sample will force the magnetic moment to align with the field direction, which results in a ferromagnetic state in high fields. Paddison $et~al.$\cite{np13_117} carried out INS measurements under a 7.8-T field applied along the $c$ axis and observed the spin-wave spectra resulting from the ferromagnetic state as shown in Fig.~\ref{Yb2} (b). In addition, there are still broad and continuum-like excitations, which are believed to be due to the disorder effect. By fitting the spectra with an anisotropic Heisenberg model plus the Zeeman term to include the field effect, they concluded that the ground state of \ymgo was a QSL in which next-nearest-neighbor exchange interactions, anisotropy, and disorder all played important roles. In \yzgo, the INS results are also quite similar\cite{Ma2017Quantum}.

While results above are all consistent with \ymgo and \yzgo being QSLs, there are several issues that make this conclusion questionable: i) The small value of the exchange interaction ($\sim$0.15~meV) restricts typical INS measurements in the overdamped regime of the excitation spectra\cite{sr5_16419,prl115_167203,np13_117,nature540_559}, so it is not clear whether the broad features both in momentum and energy are intrinsic or due to the large probing energy; ii) Since Mg$^{2+}$/Zn$^{2+}$ and Ga$^{3+}$ in the nonmagnetic layers are randomly distributed\cite{sr5_16419,prl115_167203,nature540_534}, the disorder effect, which is considered to be detrimental to the QSL phase, is severe\cite{PhysRevLett.119.157201,PhysRevLett.118.107202,np13_117};
iii) Thermal conductivity ($\kappa$) results on both \ymgo\cite{PhysRevLett.117.267202} and \yzgo\cite{Ma2017Quantum} as shown in Fig.~\ref{Yb3} strongly challenge the idea of these materials being QSLs, as we discuss in the following.

As shown in Fig.~\ref{Yb3}, in zero field, $\kappa$ is only half of that of the nonmagnetic reference sample LuMgGaO$_4$, in which only phonons contribute to $\kappa$. By fitting the zero-field data with $\kappa/T=\kappa_0/T+nT^{\beta-1}$ where $\kappa_0$ and $nT^\beta$ represent non-phonon and phonon contributions, respectively, it is shown that $\kappa_0/T$ for both \ymgo and \yzgo are effectively zero within experimental errors, similar to the case of LuMgGaO$_4$\cite{PhysRevLett.117.267202,Ma2017Quantum}. Thermal conductivity can be written as $\kappa=1/3C_{\rm m}v_{\rm F}l$, where $C_{\rm m}$, $v_{\rm F}$ and $l$ are the specific heat, Fermi velocity and the mean-free path of the quasiparticles, respectively. By assuming that the $\kappa$ at 0.1~K is totally contributed by the magnetic excitations, Xu \et estimated $l$ to be 8.6~\AA{} for \ymgo, only about 2.5 times of the spin-spin distance\cite{PhysRevLett.117.267202}. In contrast, another QSL candidate EtMe$_3$Sb[Pd$(dmit)_2]_2$ has a high $\kappa_0/T=0.2$~W\,K$^{-2}$\,m$^{-1}$, which is considered to be evidence for the presence of highly mobile quasiparticles with $l$ of $\sim$1000 times of the spin-spin distance\cite{Yamashita1246}.

Furthermore, in both \ymgo and \yzgo, it appears that the role of the magnetic excitations is to scatter off phonons that conduct heat. Therefore, when the magnetic excitations are present, $\kappa$ is reduced---this explains the the reduction of $\kappa$ as compared to that of LuMgGaO$_4$\cite{PhysRevLett.117.267202,Ma2017Quantum}. This is further manifested in the magnetic-field measurements: in a field of 9~T that opens a gap of 8.26 and 6.18~K in \ymgo\cite{PhysRevLett.117.267202} and \yzgo\cite{Ma2017Quantum}, respectively, there are almost no magnetic excitations to scatter phonons, so $\kappa$ increases. Therefore, a gapless QSL does not seem to be an applicable description for \ymgo and \yzgo, because their significant magnetic excitations, as evidenced by the large magnetic specific heat, should contribute to $\kappa$\cite{PhysRevB.76.235124,PhysRevB.72.045105}. These results, however, can be understood within a disordered-magnet picture, in which the mean-free path of the magnons is reduced with disorder, and they are not expected to conduct heat. We will discuss it further in Sec.~\ref{discussion}.

\subsection{$\kappa$-(BEDT-TTF)$_2$Cu$_2$(CN)$_3$ \label{kappacu}}

Layered organic $\kappa$-(BEDT-TTF)$_2$X are Mott insulators having two-dimensional triangular lattice as illustrated in Fig.~\ref{qslstructures}(a). Here, BEDT-TTF (ET) denotes the electron donating molecule and X represents a variety of anions with closed shells\cite{doi:10.1143/JPSJ.64.2726,Shimizu2005Spin,doi:10.1143/JPSJ.65.1340}. Among these, $\kappa$-(ET)$_2$Cu$_2$(CN)$_3$ is suggested to be a QSL\cite{PhysRevLett.91.107001,PhysRevLett.95.177001,Ohira2006,nc9_307}. The temperature dependence of magnetic susceptibility features a broad peak which can be reproduced by the triangular-lattice Heisenberg model with an exchange interaction of $J\sim250$~K\cite{PhysRevLett.91.107001}. The NMR spectra show neither a distinct broadening nor splitting down to 32~mK, suggesting the absence of magnetic order down to the temperature of 4 orders of magnitude lower than $J$\cite{doi:10.1021/cr0306541}. A $\mu$SR experiment also confirms this conclusion\cite{nature471_612}.

\begin{figure}[htb]
\centerline{\includegraphics[width=0.98\linewidth]{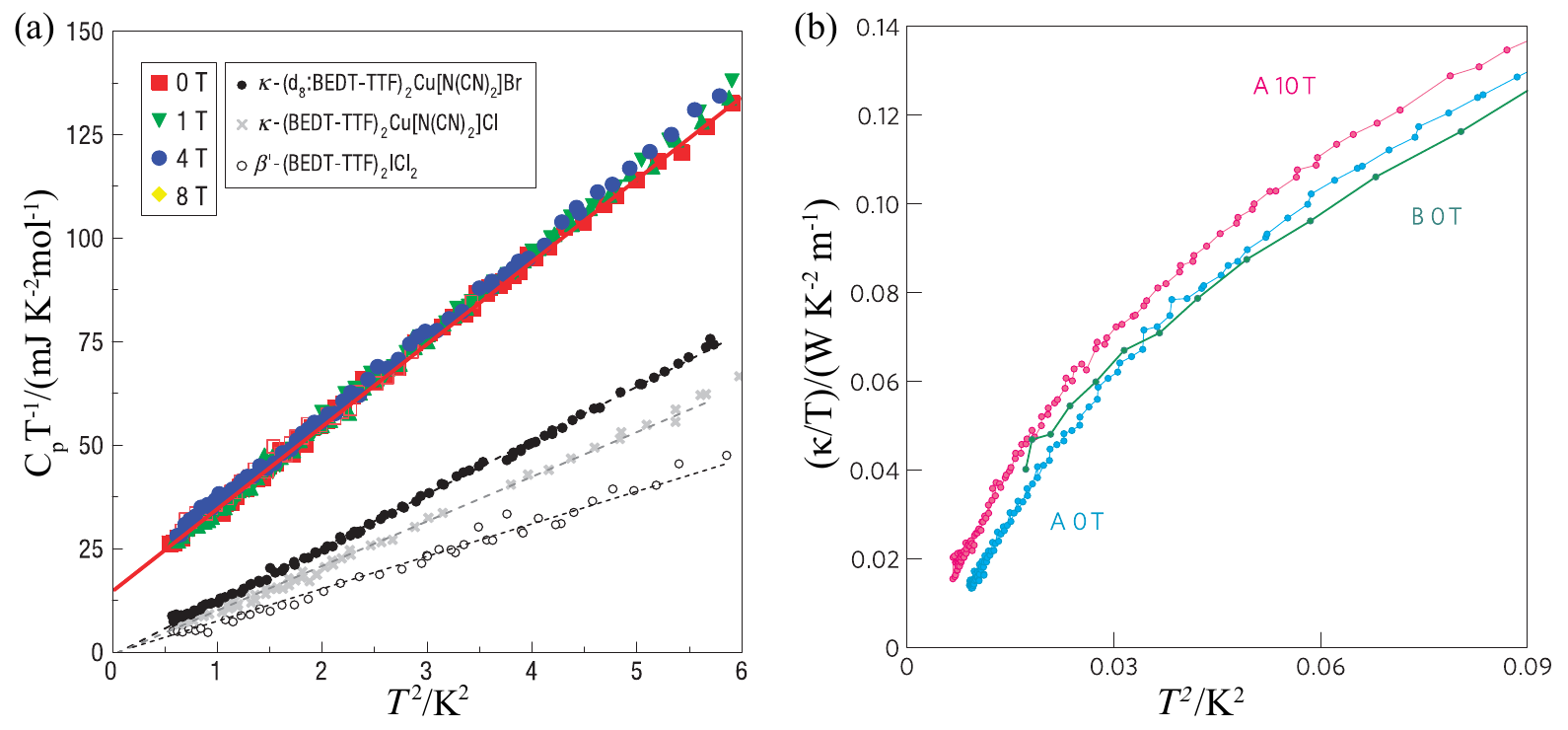}}
\caption{(a) Specific heat of $\kappa$-(ET)$_2$Cu$_2$(CN)$_3$ under different fields up to 8~T in comparison with those of other ET-based salts. From ref.~\cite{np4_459}. (b) Low-temperature thermal conductivity $\kappa$ of $\kappa$-(ET)$_2$Cu$_2$(CN)$_3$ (samples A and B). Sample A was investigated under a 10-T field applied along the $c$ axis. From ref.~\cite{np5_44}.}
\label{2-2}
\end{figure}

In Fig.~\ref{2-2}(a), temperature dependences of the specific heat for $\kappa$-(ET)$_2$Cu$_2$(CN)$_3$ under various fields are shown together with those of other ET salts\cite{np4_459}. No discernible field effect is observed. From a linear extrapolation of the data in Fig.~\ref{2-2}(a) down to zero temperature, a linearly temperature-dependent term, $i.e.$, the electronic coefficient  $\gamma$ in $C_{\rm p}T^{-1}=\gamma+\beta T^2$ is determined to be a finite value of $20\pm5$~mJ~K$^{-2}$~mol$^{-1}$, indicating the presence of gapless magnetic excitations in $\kappa$-(ET)$_2$Cu$_2$(CN)$_3$ at zero temperature\cite{np4_459}. This result is consistent with Anderson's proposal for the spin-liquid state in two-dimensional materials with the triangular-lattice structure\cite{Anderson1973153}. For other ET salts shown in Fig.~\ref{2-2}(a), $\gamma$ is zero as expected for nonmagnetic insulators. The $\gamma$ term is considered to be proportional to the spinon density of states, almost field independent\cite{np4_459}.

Thermal conductivity measurements can provide key information on the elementary excitations from the magnetic ground state in a QSL candidate, as $\kappa$ is sensitive to itinerant excitations such as the spinons. In Fig.~\ref{2-2}(b), $\kappa/T$ versus $T^2$ for $\kappa$-(ET)$_2$Cu$_2$(CN)$_3$ is shown\cite{np5_44}. The $\kappa/T$ has a extremely small value and tends to vanish as temperature decreases to zero. It should be stressed that the vanishing value of  $\kappa/T$ at $T=0$ immediately indicates the absence of low-lying fermionic excitations\cite{np5_44}, in sharp contrast to the specific heat data suggesting the presence of gapless excitations in ref.~\cite{np4_459}. Moreover, the behavior under a magnetic field up to 10~T perpendicular to the basal plane shows a nearly parallel and small shift from that under zero field. As the zero-field data, the 10-T results also suggest the absence of magnetic excitations at low temperatures. The obvious contradiction of the specific heat and thermal conductivity results leaves the nature of the magnetic ground state of $\kappa$-(ET)$_2$Cu$_2$(CN)$_3$ an open question, which will be discussed further in Sec.~\ref{discussion}.

\subsection{EtMe$_3$Sb[Pd(dmit)$_2$]$_2$}

EtMe$_3$Sb[Pd(dmit)$_2$]$_2$ is a member of the A[Pd(dmit)$_2$]$_2$ family, a layered system composed of insulating A and conducting Pd(dmit)$_2$ layers. Here, A$^+$ = Et$_x$Me$_{4-x}$Z$^+$ (Et=C$_2$H$_5$, Me=CH$_3$, Z = N, P, As, and $x$ = 0, 1, 2), and dmit is a 1,3-dithiole-2-thione-4,5-dithiolate\cite{doi:10.1246/bcsj.20130290,PhysRevLett.99.256403}. Spins are on the two-dimensional triangular lattice formed by the Pd(dmit)$_2$ molecules. Although the magnetic susceptibility shows a broad peak around 50~K, no anomaly indicative of magnetic order is observed down to 2~K\cite{doi:10.1246/bcsj.20130290,arcmp2_167}. $^{13}$C NMR measurements down to 20~mK also indicate the absence of long-range magnetic order at this temperature\cite{Itou2010Instability}.

\begin{figure}[htb]
\centerline{\includegraphics[width=0.98\linewidth]{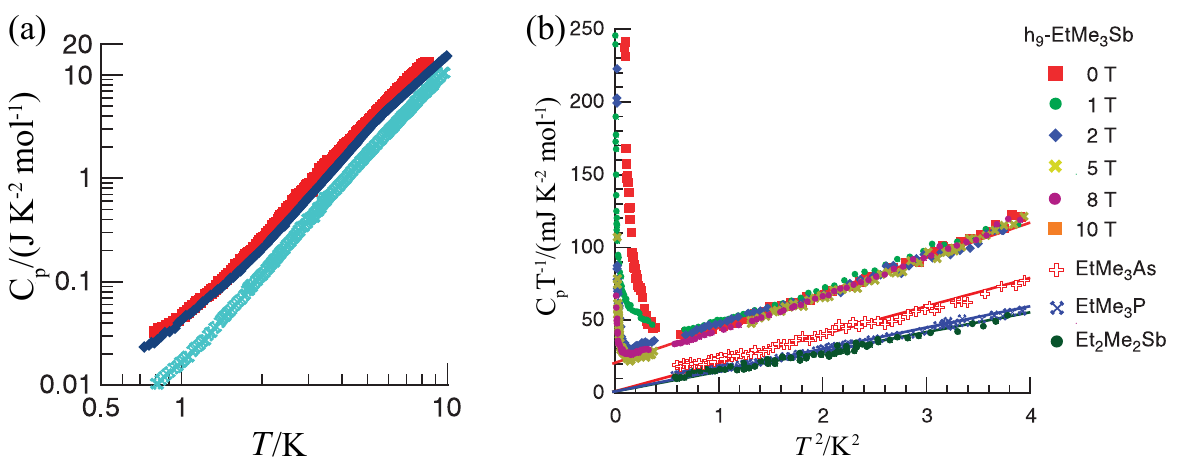}}
\caption{(a) Specific heat data of EtMe$_3$Sb[Pd(dmit)$_2$]$_2$ (red squares) plotted on a logarithmic scale. The data of EtMe$_3$P[Pd(dmit)$_2$]$_2$ (cyan crosses) and $\kappa$-(ET)$_2$Cu$_2$(CN)$_3$ (blue diamonds) are also plotted for comparison. (b) Low-temperature specific heat of EtMe$_3$Sb[Pd(dmit)$_2$]$_2$ under different fields up to 10~T. The data for other nonmagnetic systems are plotted together for comparison. From ref.~\cite{nc2_275}.}
\label{3-1}
\end{figure}

Figure~\ref{3-1}(a) shows the specific heat of EtMe$_3$Sb[Pd(dmit)$_2$]$_2$ in comparison with the results of $\kappa$-(ET)$_2$Cu$_2$(CN)$_3$\cite{np4_459} and EtMe$_3$P[Pd(dmit)$_2$]$_2$. As in the QSL candidate $\kappa$-(ET)$_2$Cu$_2$(CN)$_3$, there is no sharp peak indicative of long-range order over the entire temperature range measured. The specific heat of the QSL candidates EtMe$_3$Sb[Pd(dmit)$_2$]$_2$ and $\kappa$-(ET)$_2$Cu$_2$(CN)$_3$ are larger than that of EtMe$_3$P[Pd(dmit)$_2$]$_2$ with the nonmagnetic ground state of a valence-bond-solid type order occurring at 25~K\cite{Shimizu2007Reentrant,doi:10.1143/JPSJ.75.093701,PhysRevLett.99.256403}. This suggests that the magnetic entropy survives at low temperatures owing to the fluctuations of correlated spins in the QSL candidates\cite{nc2_275}. The low-temperature specific heat for EtMe$_3$Sb[Pd(dmit)$_2$]$_2$ is plotted as $C_{\rm p}T^{-1}$ {\it vs} $T^2$ in comparison with those of other compounds in Fig.~\ref{3-1}(b). The finite electronic specific heat coefficient $\gamma=19.9$~mJ~K$^{-2}$~mol$^{-1}$ suggests the presence of gapless excitations at zero temperature\cite{nc2_275}. The value of $\gamma$ is almost field independent, consistent with the presence of a spinon Fermi surface\cite{nc2_275}. On the other hand, the non-QSL candidates have $\gamma$ that is effectively zero.

\begin{figure}[htb]
\centerline{\includegraphics[width=0.98\linewidth]{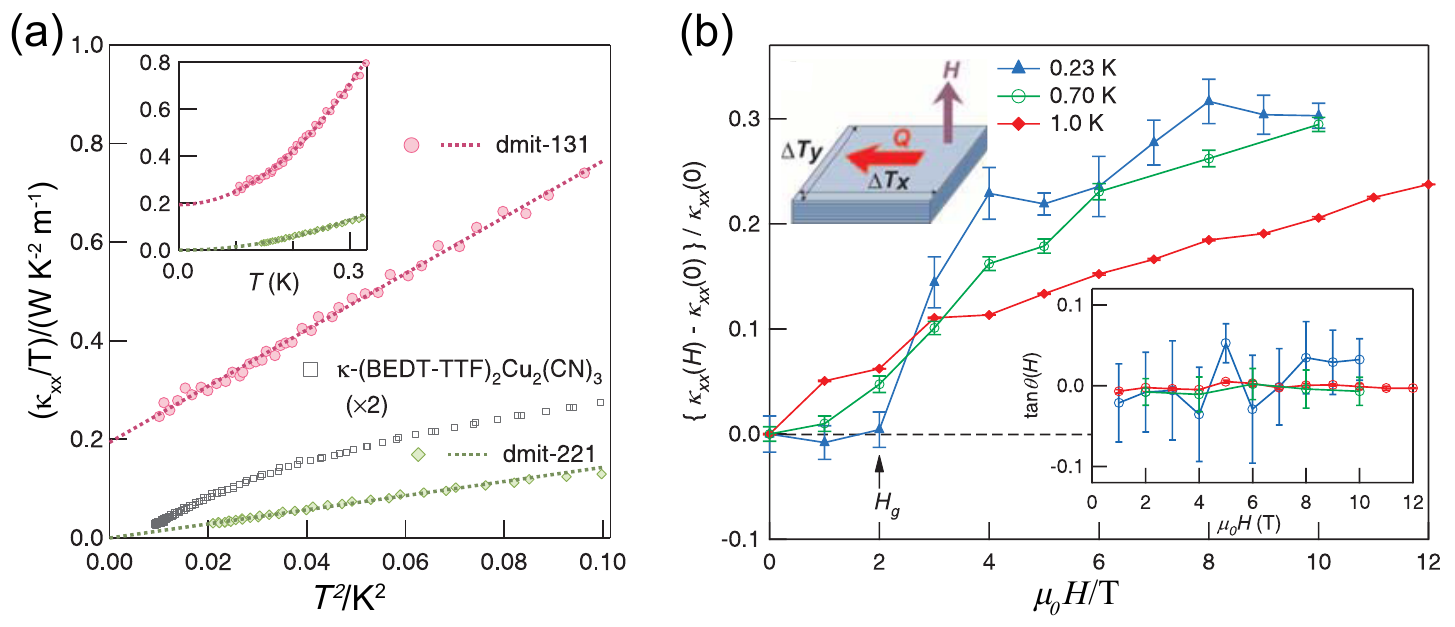}}
\caption{(a) Low-temperature plot of $\kappa_{xx}/T$ as a function of $T^2$ of dmit-131 (circles), dmit-221 (diamonds), and $\kappa$-(ET)$_2$Cu$_2$(CN)$_3$ (squares). The inset shows $\kappa/T$ plotted as a function of $T$. (b) Field dependence of $\kappa$ normalized by the zero-field value, [$\kappa_{xx}$($H$)- $\kappa_{xx}$(0)]/$\kappa_{xx}$(0) of dmit-131 at low temperatures. Upper left inset shows a schematic of the measurement setup. The lower left inset shows thermal-Hall angle tan$\theta(H)=\kappa_{xy}$/($\kappa_{xx} - \kappa_{xx}^{\rm ph}$) as a function of the field value $\mu_0H$ at 0.23~K (triangles), 0.70~K (circles), and 1.0~K (diamonds). $\kappa_{xy}$ and $\kappa_{xx}^{\rm ph}$ are the thermal-Hall conductivity and phonon-contributed thermal conductivity, respectively. From ref.~\cite{Yamashita1246}.}
\label{3-2}
\end{figure}

In Fig.~\ref{3-2}(a), the longitudinal thermal conductivity $\kappa_{xx}$ over $T$ of EtMe$_3$Sb[Pd(dmit)$_2$]$_2$ is plotted against $T^2$ in comparison with those of another QSL candidate $\kappa$-(ET)$_2$Cu$_2$(CN)$_3$ and a nonmagnetic compound Et$_2$Me$_2$Sb[Pd(dmit)$_2$]$_2$\cite{Yamashita1246}. Interestingly, in contrast to the latter two, where $\kappa_{xx}/T$ is 0 as $T$ approaches 0, the residual $\kappa_{xx}/T$ for EtMe$_3$Sb[Pd(dmit)$_2$]$_2$ is 0.2~W\,K$^{-2}$\,m$^{-1}$\cite{Yamashita1246}. The presence of large residual $\kappa_{xx}$ is also confirmed in the inset of Fig.~\ref{3-2}(a) in which $\kappa_{xx}/T$ is plotted as a function of $T$. This result is consistent with the observation of gapless magnetic excitations in the specific heat measurement. By using $\kappa_{xx}=1/3C_{\rm m}v_{\rm F}l$,  it is suggested that there are highly mobilized magnetic excitations with a mean-free path $l$ of $\sim$1000 times of the spin-spin distance\cite{Yamashita1246}. Remarkably, this is a rare example where thermal conductivity measurements show a finite $\kappa_{xx}/T$ at zero temperature. The results indicate the presence of gapless magnetic excitations consistent with the specific heat data\cite{nc2_275}.

The magnetic-field dependence of $\kappa_{xx}$ is shown in Fig.~\ref{3-2}(b)\cite{Yamashita1246}. At the lowest temperature, $\kappa_{xx}$($H$) under low fields is insensitive to the magnetic field strength $\mu_0H$ but displays a steep increase above a characteristic magnetic field of 2~T. This behavior is less profound with increasing temperature. At 1~K, $\kappa_{xx}$ becomes linear. The observed field dependence is interpreted as the presence of spin-gap-like excitations at low temperatures, along with the gapless excitations inferred from the residual $\kappa_{xx}/T$ and finite $\gamma$\cite{Yamashita1246,nc2_275}. The gap behavior is also suggested from the zero-field thermal conductivity measurement where $\kappa_{xx}/T$ $vs.$ $T$ shows a broad peak at $\sim$1~K\cite{Yamashita1246}.

\subsection{ZnCu$_3$(OH)$_6$Cl$_2$}

\begin{figure}[htb]
\centerline{\includegraphics[width=0.8\linewidth]{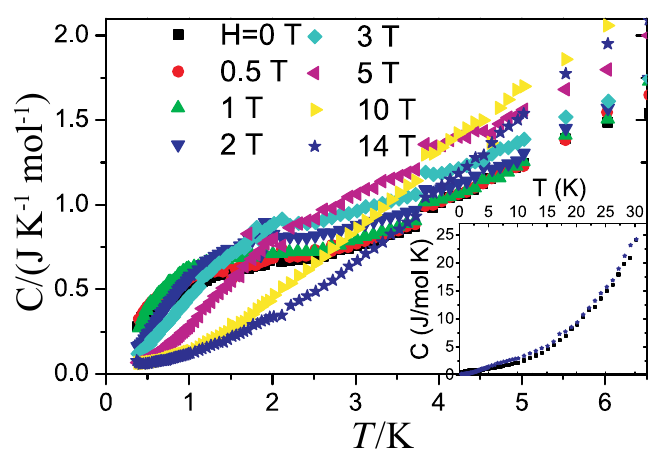}}
\caption{Specific heat of Herbertsmithite measured under several different magnetic fields up to 14 T. Inset: Specific heat plotted over a wider temperature range under zero (squares) and 14-T (stars) fields. From ref.~\cite{prl98_107204}.}
\label{4-2}
\end{figure}

ZnCu$_3$(OH)$_6$Cl$_2$ known as Herbertsmithite has a three-dimensional rhombohedral structure and consists of two-dimensional kagom\'{e}-lattice planes of spin-1/2 Cu$^{2+}$ ions separated by nonmagnetic layers formed by Zn$^{2+}$\cite{Nytko2008A}. Geometrical frustration on such  a structurally perfect kagom\'{e} lattice is expected to be strong and many interesting phenomena may emerge\cite{doi:10.1021/ja053891p,RevModPhys.88.041002}. Because large-size single crystals are available for this material\cite{PhysRevB.83.100402}, it has been heavily studied by various experimental techniques. Previous measurement results\cite{prl98_107204,PhysRevLett.98.077204,PhysRevLett.103.237201} of polycrystalline Herbertsmithite suggest that there is no static magnetic order nor spin freezing down to 50~mK despite the large Curie-Weiss temperature of -314~K\cite{Nytko2008A}. Specific heat shows no sharp $\lambda$-type peak down to $\sim$100~mK\cite{prl98_107204}. When applying a magnetic field to the sample, the specific heat is changed rapidly as shown in Fig.~\ref{4-2}~(ref.~\cite{prl98_107204}). Under a 14-T field, the low-temperature specific heat is largely reduced, because magnetic excitations are gapped, leaving only phonons contributing to the specific heat. On the other hand, at high temperatures where phonons dominate the specific heat, there is hardly any field effect.

\begin{figure}[htb]
\centerline{\includegraphics[width=0.9\linewidth]{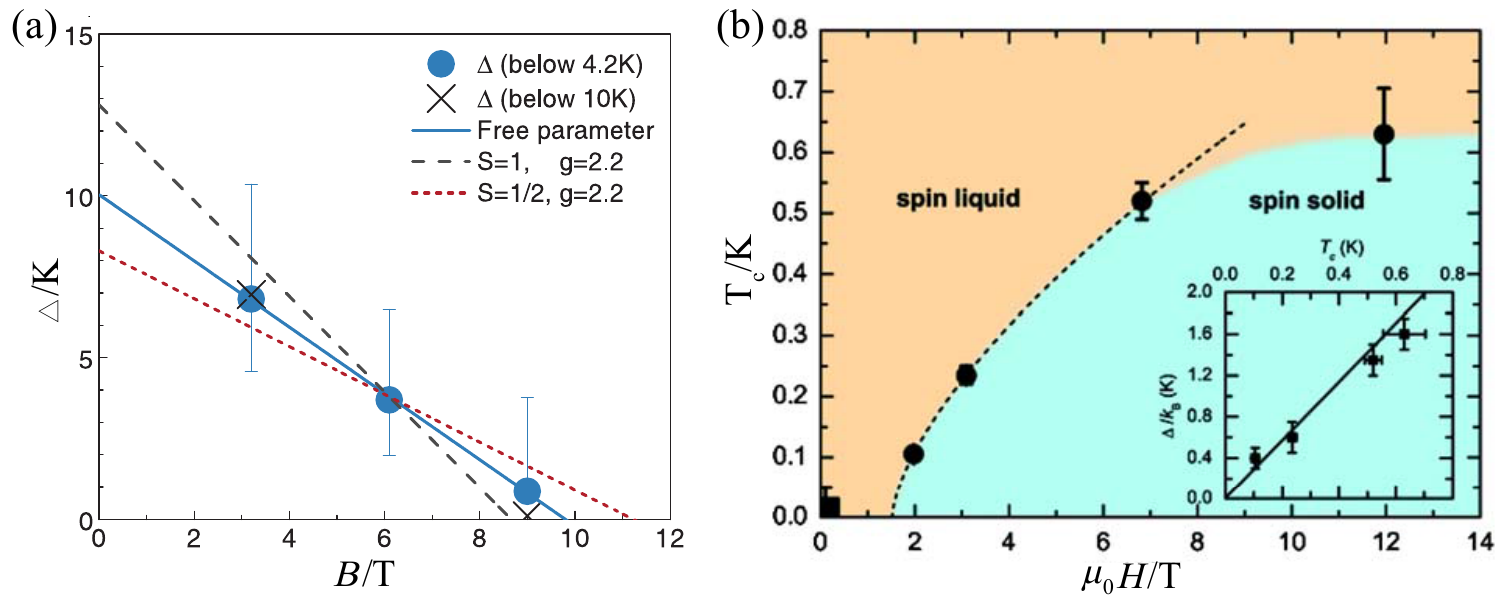}}
\caption{(a) Spin excitation gap $\Delta$ of Herbertsmithite obtained by fitting NMR spectra with a temperature range up to 4.2~K (circles) and 10~K (crosses). Dashed and dotted lines represent the best fits under the constraint of $S = 1$ and 1/2, respectively. The solid line represents the best free-parameter fit. From ref.~\cite{science350_655}. (b) Transition temperature $T_c$ versus magnetic field phase diagram of Herbertsmithite. The inset plots $\Delta$ as a function of $T_c$. From ref.~\cite{PhysRevLett.107.237201}.}
\label{4-4}
\end{figure}

The availability of large single crystals of Herbertsmithite\cite{PhysRevB.83.100402} makes it possible to carry out INS measurements, which reveal a broad continuum expected for a QSL state\cite{nature492_406}. The INS results show no spin gap down to 0.25~meV. However, previous calculation results by density matrix renormalizaton group indicate that the ground state of the Heisenberg model on a kagom\'{e} lattice is a fully gapped QSL\cite{Yan1173}. To understand the magnetic ground state of Herbertsmithite, Asaba $et \; al.$\cite{PhysRevB.90.064417} performed magnetization measurements on a single crystal using torque magnetometry with an intense magnetic field up to 31~T. Based on the observation that the effective magnetic susceptibility under high fields is independent of the temperature within the low-temperature range between 20~mK and 5~K, they considered the ground state to be gapless, consistent with the INS results\cite{nature492_406}.

However, as shown in Fig.~\ref{4-4}(a), NMR measurements on single crystals of Herbertsmithite reveal a finite gap value $0.86\pm0.26$~meV by extrapolating the fitted values of gap to zero field\cite{science350_655}.
This value is close to a later INS measurement which identifies a spin gap of 0.7~meV by modeling the momentum-integrated data as the sum of the contributions from a damped harmonic oscillator and the gapped excitations\cite{PhysRevB.94.060409}. We note that a more recent calculation employing the tensor network shows that the ground state is a gapless QSL\cite{PhysRevLett.118.137202}. Concerning the nature of the ground state for this kagom\'{e} compound, it turns out that no consensus has been reached so far.

Moreover, NMR measurements on polycrystalline samples identify a quantum critical point at a critical field $B_c$ of 1.53~T, at which the system is believed to evolve from a spin liquid to a solid, as illustrated in Fig.~\ref{4-4}(b)\cite{PhysRevLett.107.237201}. The high-field, low-temperature spin-solid state is featured by frozen spins with slow relaxations. As shown in the inset of Fig.~\ref{4-4}(b), the gap size in the spin-solid state is roughly proportional to the critical temperature $T_c$\cite{PhysRevLett.107.237201}.

\section{Kitaev materials---$\alpha$-R\lowercase{u}C\lowercase{l}$_3$}

Different from aforementioned QSL candidates which have either triangular or kagom\'{e} lattice where antiferromagnetic exchange interactions are geometrically frustrated, the Kitaev QSL has honeycomb lattice where the frustration on a single site arises from the bond-dependent spin anisotropy\cite{Kitaev2006Anyons}. Possible realization of such an exotic state has been suggested in SOC-assisted Mott insulators such as Na$_2$IrO$_3$\cite{prl102_017205,prl105_027204,PhysRevLett.102.256403,prl110_097204,PhysRevB.93.174425,PhysRevB.95.144406,0953-8984-29-49-493002}, and Li$_2$IrO$_3$\cite{nc5_4203,PhysRevLett.114.077202,PhysRevB.93.195158,PhysRevB.95.144406,nc7_12286}. These materials have the honeycomb lattice as shown in Fig.~\ref{qslstructures}(c). Due to the combination of the cubic crystal electric field, strong SOC, and electronic correlations, the ground state is Krammers doublets with an effective spin of 1/2~(refs~\cite{PhysRevLett.110.076402,PhysRevLett.109.266406,PhysRevB.82.064412}). Furthermore, because of the strong SOC, the effective spin is expected to be anisotropic due to the spatial anisotropy of the 5$d$ orbitals of Ir$^{4+}$. Therefore, the bond-dependent anisotropic Kitaev interaction may be realized on the honeycomb lattice\cite{Kitaev2006Anyons}. However, it is found that they are not QSLs but are instead magnetically ordered at low temperatures\cite{PhysRevB.82.064412,PhysRevB.83.220403,PhysRevLett.108.127204,PhysRevB.85.180403,nc5_4203,PhysRevLett.114.077202,PhysRevB.93.195158,PhysRevB.95.144406,nc7_12286,0953-8984-29-49-493002}. Nevertheless, subsequent experimental and theoretical works suggest that the magnetic orders in these materials are unconventional, signifying the presence of notable Kitaev interaction. For instance, it is suggested that the zigzag magnetic order in Na$_2$IrO$_3$ can be understood within a Heisenberg-Kitaev model\cite{PhysRevLett.108.127203,prl110_097204}. In the phase diagram constructed using this model, there are regions where the Heisenberg interaction is small and the Kitaev interaction is dominant, leading to the Kitaev QSL phase. Because of this, finding Kitaev QSLs in related materials is still encouraging. In the following, we will discuss another Kitaev material $\alpha$-RuCl$_3$, which has been the focus of recent research due to the availability of high-quality single crystals and feasibility of neutron scattering measurements.

$\alpha$-RuCl$_3$ has two-dimensional honeycomb layers formed by the 4$d$ Ru$^{3+}$ ions\cite{nature199_1089,J19670001038,PhysRevB.90.041112,PhysRevB.91.094422,PhysRevB.91.144420,PhysRevB.93.134423}.
In fact, realization of the Kitaev interaction in materials with 4$d$ electrons does not sound very promising in the beginning, because their SOCs are smaller compared to those of the 5$d$ systems. However, although the absolute value of the SOC in RuCl$_3$ is smaller, the almost-90$^\circ$ bond angles of the Cl-Ru-Cl bonds of the edge-shared RuCl$_6$ octahedra makes the cubic crystal electric field win and the SOC become a dominant effect\cite{nature199_1089,J19670001038,PhysRevB.90.041112,PhysRevB.91.094422,PhysRevB.91.144420,PhysRevB.93.134423,GUIZZETTI197934,pss44_245,PhysRevB.53.12769,PhysRevB.50.2095}. Thus, similar to iridates, \rucl is also an SOC-assisted Mott insulator with an effective spin of 1/2, and the strong spatial anisotropy of the 4$d$ orbitals combined with the SOC makes the bond-dependent Kitaev interaction significant\cite{Kitaev2006Anyons,nature199_1089,J19670001038,PhysRevB.96.054410}.  However, similar to Na$_2$IrO$_3$, the ground state of \rucl is not a Kitaev QSL, but a zigzag magnetic order state instead\cite{nature199_1089,J19670001038,PhysRevB.90.041112,PhysRevB.91.094422,PhysRevB.91.144420,PhysRevB.93.134423}.
It has been proposed that the zigzag order is an indication for the presence of the Kitaev interaction in this system\cite{PhysRevB.90.041112,PhysRevB.91.094422,PhysRevB.91.144420,PhysRevB.93.134423,prl102_017205,prl105_027204}. Moreover, INS results indicate that the ground state of \rucl may be proximate to the Kitaev QSL phase\cite{NM}, and both INS\cite{NM,ranprl2017,Banerjee1055,np13_1079} and Ramman studies\cite{PhysRevLett.114.147201,np12_912} observe broad continuous magnetic excitations that can be associated with fractionalized excitations resulting from the Kitaev QSL phase.

\begin{figure}[htb]
\centering
\includegraphics[width=0.7\linewidth,trim=0mm 0mm 0mm 0mm]{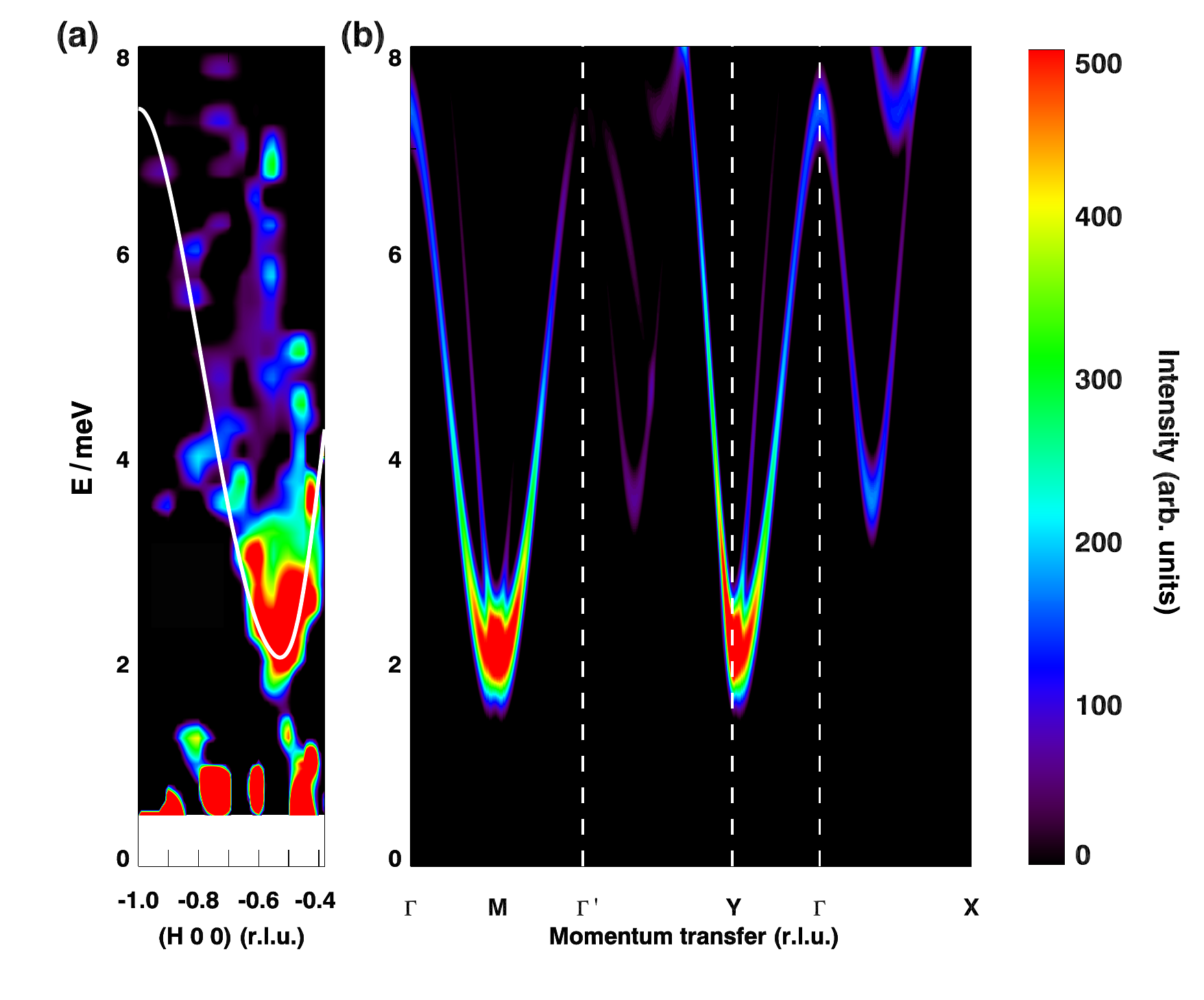}\caption{\label{spinwaves}{(a) Magnetic dispersion along the [100] direction obtained by INS experiments. The solid line is the calculated dispersion as shown in (b). (b) Calculated spin-wave spectra along high-symmetry paths. From ref.~\cite{ranprl2017}.}}
\end{figure}

By analyzing the INS spectra, magnetic interactions governing the ground state can be extracted. In Fig.~\ref{spinwaves}(a), we show the spin-wave excitation spectra resulting from the zigzag order state\cite{ranprl2017}. Instead of using the widely considered Heisenberg-Kitaev ($H$-$K$) model\cite{prl102_017205,prl105_027204} to fit the spectra, a $K$-$\Gamma$ effective-spin model is used, where $K$ and $\Gamma$ represent Kitaev and off-diagonal exchange interactions, respectively. This minimal model describing the ground state of \rucl is first proposed by Wang \et in ref.~\cite{PhysRevB.96.115103}. They recognize the importance of the off-diagonal interactions, but on the other hand, they find that the Heisenberg interactions are at least an order of magnitude smaller than either $K$ or $\Gamma$ in the parameter range relevant to this material. Fits to the INS spectra with this model yield a ferromagnetic $K$ of -6.8~meV and a $\Gamma$ of 9.5~meV\cite{ranprl2017}, very close to the calculated results of $K=-5.5$~meV and $\Gamma=7.6$~meV reported in ref.~\cite{PhysRevB.96.115103}. These results unambiguously demonstrate that the Kitaev interaction is large and exists in real material.

As mentioned above, the ground state of \rucl is the zigzag order state instead of a Kitaev QSL. Nevertheless, the magnetic order is rather fragile, with an ordered moment of $\sim$0.4$\mu_B$ and an ordering temperature of $\sim$8~K\cite{nature199_1089,J19670001038,PhysRevB.90.041112,PhysRevB.91.094422,PhysRevB.91.144420,PhysRevB.93.134423,ranprl2017}. Such a fragile order can be fully suppressed by either an in-plane magnetic field\cite{PhysRevB.91.094422,PhysRevB.91.180401,PhysRevB.92.235119} or pressure\cite{PhysRevB.96.205147}. How do the magnetic excitations behave in the high-field state? Is the high-field disordered state a QSL? If the high-field state is a QSL, what is the relationship between this phase and the long-sought Kitaev QSL? To answer these questions, measurements utilizing various experimental techniques have been carried out\cite{PhysRevB.91.094422,PhysRevB.91.180401,PhysRevB.92.235119,kjNMR,PhysRevB.96.041405,PhysRevB.95.180411,PhysRevLett.118.187203,PhysRevB.95.245104,PhysRevLett.119.227201,PhysRevLett.119.227202,PhysRevB.96.241107,npjqm3_8,PhysRevLett.120.117204}. Some of these results are discussed as following.

\begin{figure}[htb]
\centering
\includegraphics[width=0.99\linewidth,trim=0mm 0mm 0mm 0mm]{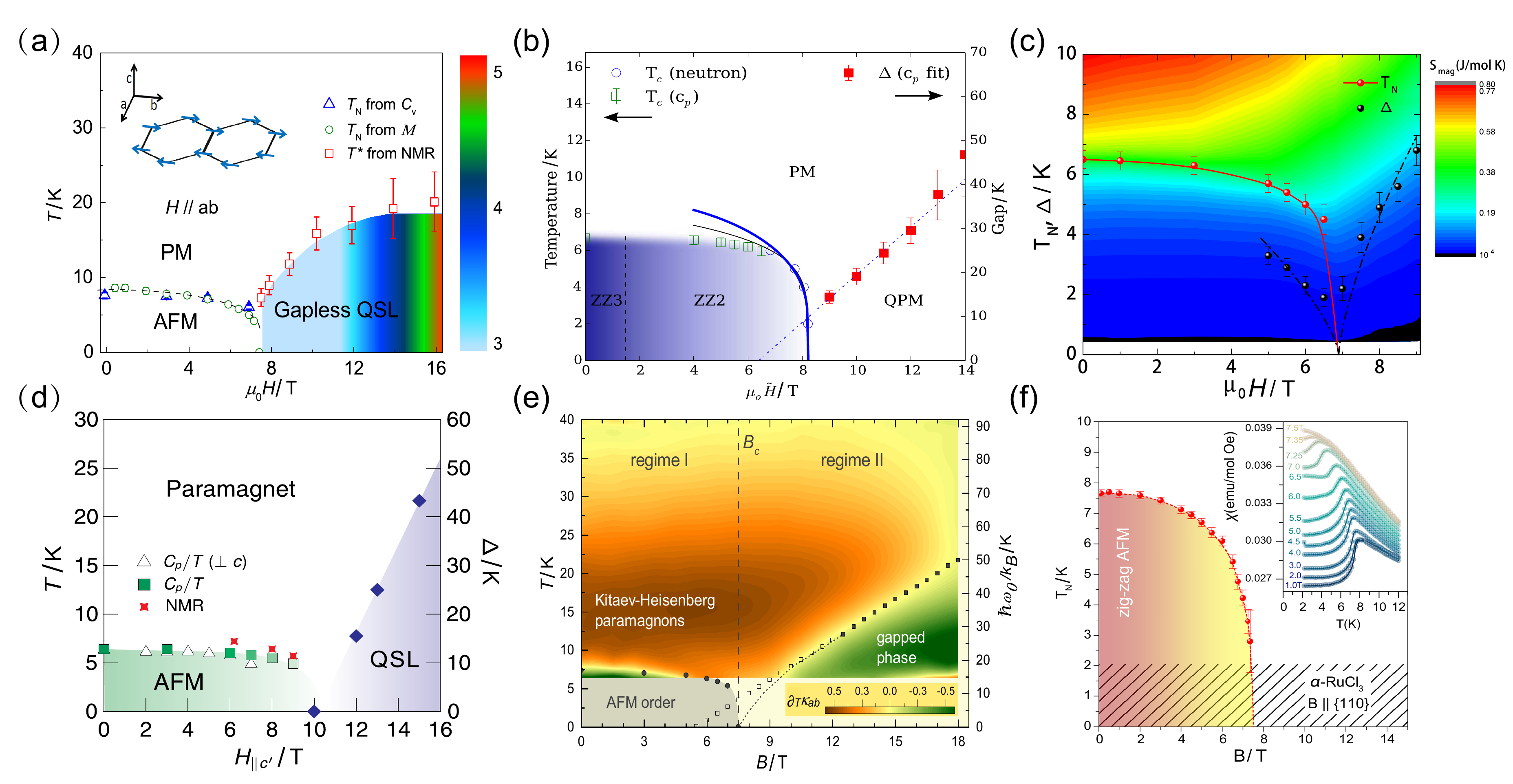}
\caption{\label{fig2}{{\bf Magnetic phase diagram of \rucl obtained from various measurements.} (a) PM and AFM represent paramagnet and zigzag order, respectively. Low--field region of the phase diagram is  constructed using magnetization and specific heat data, and the high-field region is using NMR data. The magnetic field is applied in the $a$-$b$ plane. The contour map indicates the exponent of the temperature dependence of the spin-lattice relaxation rate. The inset illustrates the zigzag order in the low-field state. From ref.~\cite{kjNMR}. (b) ZZ and QPM represent zigzag order and quantum paramagnet, respectively. The phase boundary between ZZ2 and PM is the transition temperature obtained from specific heat and neutron diffraction measurements. The thick solid line is a fit with the transverse-field Ising model, and the thin solid line is a power-law fit. The dashed line is a power-law fit to the gap size $\Delta$. From ref.~\cite{PhysRevB.95.180411} (c) Transition temperatures and gap values obtained from specific heat measurements. The solid line is a guide to the eye. Dashed lines are the fits of the gap function. The magnetic entropy is shown in a color scale. From ref.~\cite{PhysRevB.96.041405}. (d) Phase diagram obtained from specific heat and NMR measurements along with the field dependence of the spin gap $\Delta$ extracted from the nuclear-spin relaxation rate (right axis). From ref.~\cite{PhysRevLett.119.037201}. (e) False-color representation of the $T$ derivative of the $ab$-plane
thermal conductivity ($\kappa_{ab}$) together with the gap values
(solid squares) extracted from the phononic fits. The color scale is
in the unit of W/K$^2$\,m. From ref.~\cite{PhysRevLett.120.117204}. (f) Magnetic transition temperature as a function of field obtained from susceptibility measurements with field applied along the [110] direction. The inset shows the susceptibility data. From ref.~\cite{npjqm3_8}.}}
\end{figure}

By following the magnetic-field dependence of the magnetization and specific heat, it is found that the zigzag order is gradually suppressed, and the system becomes a magnetically disordered state at $\sim$7.5~T\cite{kjNMR}, consistent with earlier reports on the field effect\cite{PhysRevB.91.094422,PhysRevB.91.180401,PhysRevB.92.235119}. NMR spectra on high-quality single crystals also indicate that there is a quantum critical point at $B_c\sim$7.5~T\cite{kjNMR}. Above $B_c$, the spin-lattice relaxation rate $1/T_1$ of $^{35}$Cl shows a power-law behavior as $1/T_1\sim T^\alpha$. In a field range between 8 and 16~T, $\alpha\approx3$, suggesting a field-induced QSL featuring Dirac nodal-like spin excitations. A phase diagram summarizing these results is shown in Fig.~\ref{fig2}(a). Intensive research on the high-field state utilizing various techniques such as magnetization\cite{PhysRevB.91.094422,PhysRevB.91.180401,PhysRevB.92.235119,kjNMR,PhysRevLett.120.067202,npjqm3_8}, specific heat\cite{PhysRevB.91.094422,kjNMR,PhysRevB.96.041405,PhysRevB.95.180411,PhysRevLett.119.037201,PhysRevLett.120.067202}, magnetodielectric\cite{PhysRevB.95.245104}, neutron diffraction\cite{PhysRevB.92.235119,PhysRevB.95.180411,npjqm3_8}, NMR\cite{kjNMR,PhysRevLett.119.037201}, magnetic torque\cite{PhysRevLett.118.187203}, thermal conductivity\cite{PhysRevLett.118.187203,PhysRevLett.120.067202,PhysRevLett.120.117204}, terahertz spectroscopy\cite{PhysRevLett.119.227201,PhysRevLett.119.227202}, and electron spin resonance\cite{PhysRevB.96.241107}, has resulted in many somewhat similar phase diagrams, some of which are shown in Fig.~\ref{fig2}.

\begin{figure}[ht]
\centering
\includegraphics[width=0.7\linewidth,trim=0mm 0mm 0mm 0mm]{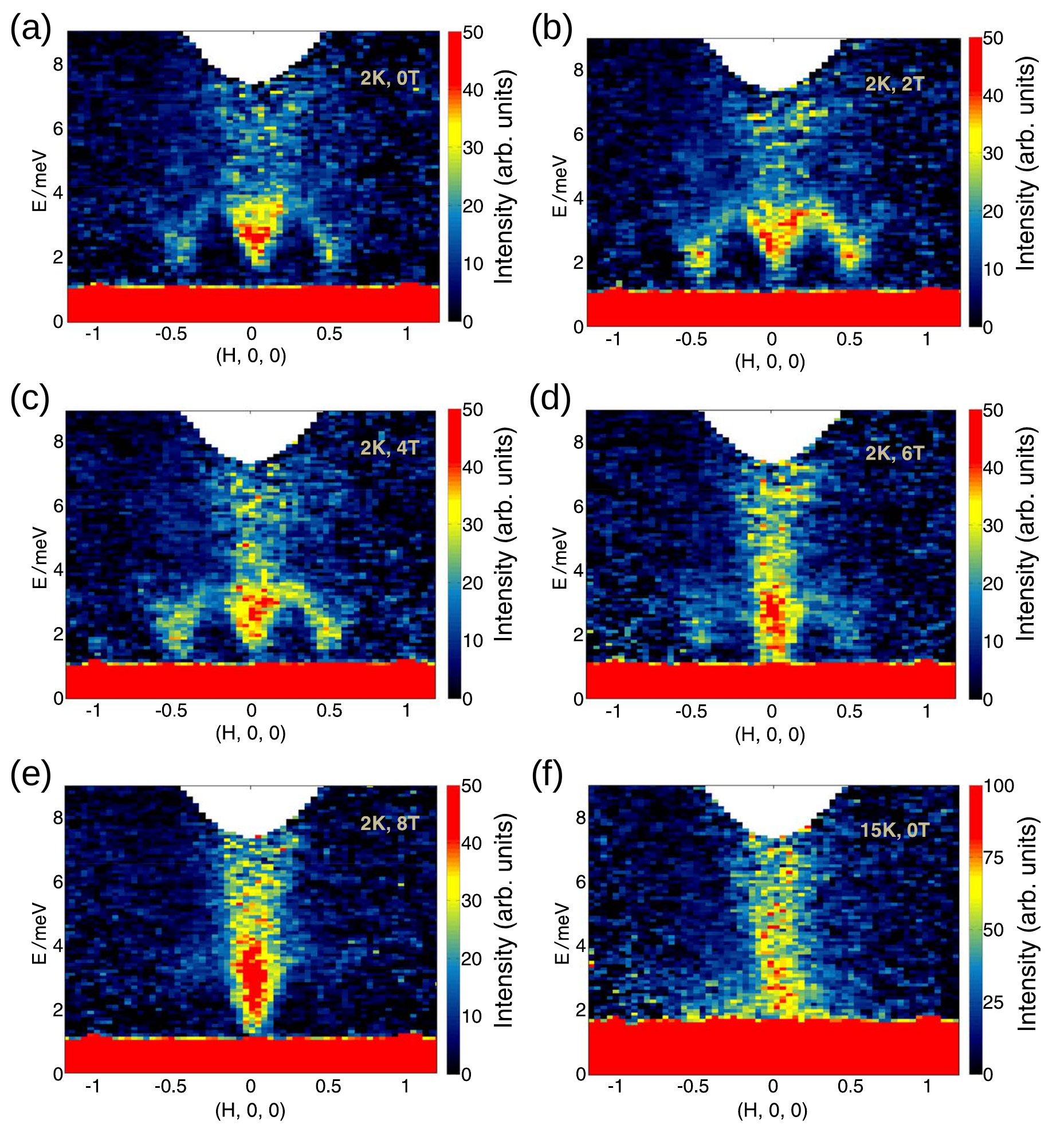}
\caption{\label{fig4npj}{(a)-(e) Magnetic-field evolution of the INS spectra along the [100] direction, measured at $T = 2$~K with an external magnetic field applied in the $a$-$b$ plane. (f) The zero-field data at 15~K, above the ordering temperature. From ref.~\cite{npjqm3_8}.}}
\end{figure}

On one hand, these phase diagrams all show that the zigzag magnetic order is gradually suppressed by an in-plane magnetic field, and the system reaches a quantum critical point around $B_c\approx7.5$~T. Furthermore, accumulating evidence suggests that the high-field disordered state above $B_c$ is a QSL. In particular, Banerjee $et~al$. have carried out INS measurements to examine the magnetic-field evolution of the magnetic excitations, and some of the results are shown in Fig.~\ref{fig4npj}\cite{npjqm3_8}. They find that the spin-wave excitations associated with the zigzag order near the M point is suppressed with the field. As shown in Fig.~\ref{fig4npj}(e), above $B_c$, at $\mu_0H=8$~T, excitations near the M point are completely gone, and only excitations near the $\Gamma$ point remain, similar to the results under zero field above the ordering temperature, as shown in Fig.~\ref{fig4npj}(f). By comparing with calculations, they suggest that the excitations under high fields resemble those of a Kitaev QSL\cite{npjqm3_8}.

\begin{figure}[htb]
\centering
\includegraphics[width=0.98\linewidth,trim=0mm 0mm 0mm 0mm]{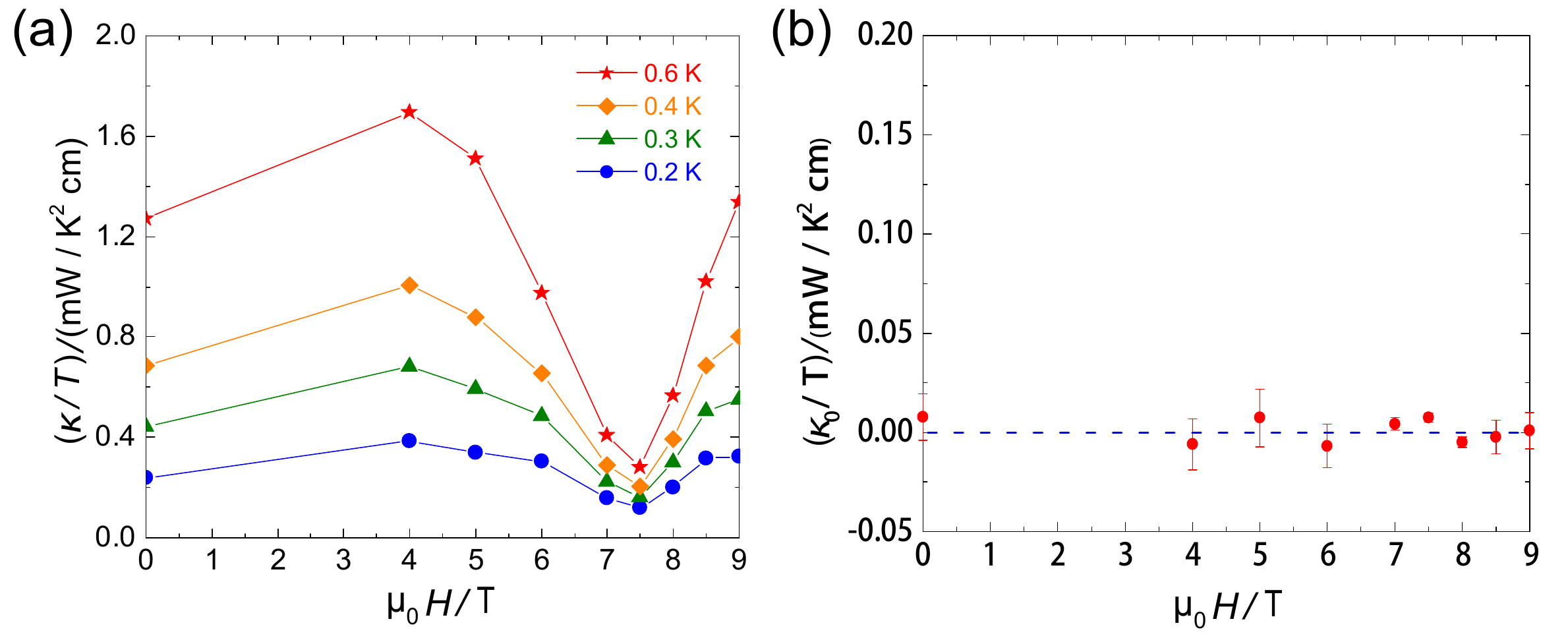}
\caption{\label{fig3}{(a) Field dependence of $\kappa$/$T$ at various temperatures. The minimum of $\kappa$/$T$ at $\sim$7.5~T corresponds to $B_c$, where the zigzag magnetic order disappears. (b) Field dependence of the residual linear term $\kappa_0$/$T$. From ref.~\cite{PhysRevLett.120.067202}.}}
\end{figure}

On the other hand, the nature of the field-induced QSL phase, in particular, whether the low-energy magnetic excitations associated with this state are finite\cite{kjNMR,PhysRevLett.118.187203} or fully gapped\cite{PhysRevB.96.041405,PhysRevB.95.180411,PhysRevLett.119.227202,PhysRevLett.119.037201,PhysRevB.96.241107,npjqm3_8}, is still under debate. To resolve this issue, Yu \et\cite{PhysRevLett.120.067202} have performed ultralow-temperature thermal conductivity measurements down to 80~mK under magnetic fields, and the results are summarized in Fig.~\ref{fig3}. As shown in Fig.~\ref{fig3}(a), $\kappa/T$ increases with field and then decreases to a minimum at the critical field $B_c$. Above $B_c$, $\kappa/T$ increases with field again. These results clearly show that there is a quantum critical point at $B_c$, consistent with other works\cite{kjNMR,PhysRevB.96.041405,PhysRevB.95.180411,PhysRevLett.119.037201,PhysRevLett.118.187203,PhysRevLett.119.227202,PhysRevB.96.241107}. By examining the residual $\kappa$ at zero temperature, it is found that $\kappa_0$ is effectively zero in the whole field range probed~[Fig.~\ref{fig3}(b)]. In the low-field range below $B_c$, there is a spin-anisotropy gap about 2~meV associated with the zigzag order\cite{NM,ranprl2017,npjqm3_8}. Therefore, absence of the thermal conductivity at zero temperature is expected. If the high-field state is also fully gapped as suggested in refs~\cite{PhysRevB.96.041405,PhysRevB.95.180411,PhysRevLett.119.227202,PhysRevLett.119.037201,PhysRevB.96.241107,npjqm3_8}, these thermal conductivity results are easily explainable. However, as shown in Fig.~\ref{fig2}(a), a gapless QSL state near $B_c$ is suggested\cite{kjNMR}. We here provide one possible solution to reconcile this discrepancy. According to the report, this gapless state is in fact featured by Dirac-like excitations with gap nodes in the momentum space\cite{kjNMR}. In this case, the magnetic density of states, represented as $C_{\rm m}$ in $\kappa=1/3C_{\rm m}v_{\rm F}l$ is small. At present, the estimated magnetic specific heat $C_{\rm m}$ has big uncertainties due to the lack of proper reference sample to subtract the phonon contributions\cite{PhysRevB.91.094422,kjNMR,PhysRevB.96.041405,PhysRevB.95.180411,PhysRevLett.119.037201,PhysRevLett.120.067202}. If the real $C_{\rm m}$ is small, then the gapless state with nodal excitations proposed in ref.~\cite{kjNMR} can also be consistent with the thermal conductivity results\cite{PhysRevLett.120.067202}.

\section{Discussions \label{discussion}}

As partially reflected from the discussions above, research on the QSL candidates has been quite dynamics. A lot of progress has been made already in recent years. However, it still lacks an ideal QSL candidate so far. Quite often, the spin-``liquid" behavior may have some other origins than quantum fluctuations. Below, we will show some examples.

There is accumulating evidence suggesting \ymgo to be a promising candidate as a gapless QSL\cite{sr5_16419,prl115_167203,np13_117,nature540_559,PhysRevLett.117.097201,PhysRevLett.118.107202,arXiv:1704.06468}.
However, the report of no positive contributions from the magnetic excitations to the thermal conductivity is difficult to be reconciled with the gapless QSL picture\cite{PhysRevLett.117.267202}. One possibility is that the severe disorder effect caused by the random mixing of Mg$^{2+}$ and Ga$^{3+}$ makes the otherwise itinerant spinons localized and thus not conduct heat\cite{sr5_16419,prl115_167203,nature540_534,PhysRevLett.118.107202,PhysRevLett.117.267202}. However, the disorder is considered to be detrimental to the QSL phase for this compound\cite{PhysRevLett.119.157201}.  Ma \et\cite{Ma2017Quantum} have carried out measurements on \yzgo, a sister compound of \ymgo, utilizing various techniques, including d.c. susceptibility, specific heat, INS, and ultralow-temperature thermal conductivity. They have found that a spin-glass phase can explain the experimental observations in \yzgo: including no long-range magnetic order, prominent broad excitation continua observed by INS, and absence of magnetic thermal conductivity. By analogy, they suggest the spin-glass phase to be also applicable to \ymgo.

\begin{figure}[htb]
\centerline{\includegraphics[width=0.98\linewidth]{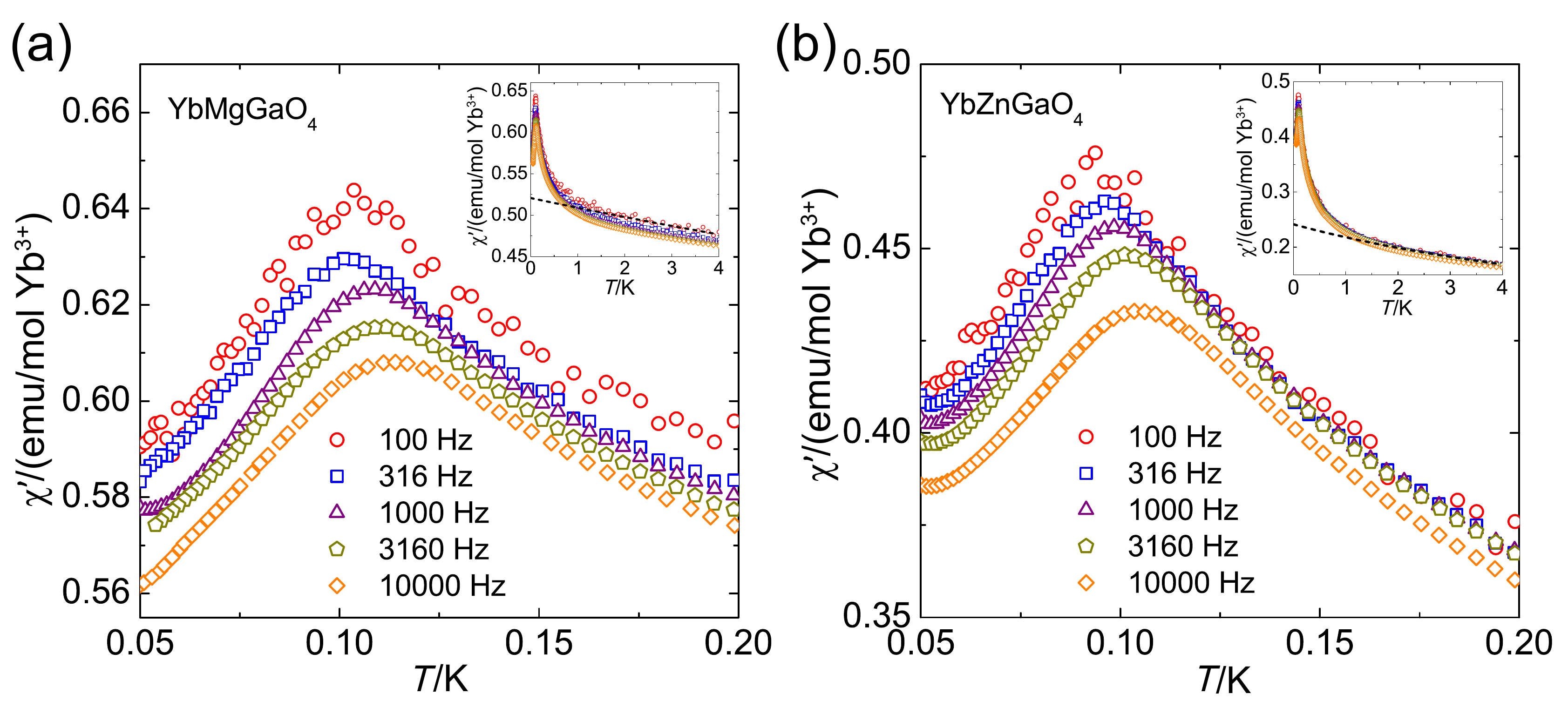}}
\caption{(a) and (b) Temperature dependence of the a.c. susceptibility ($\chi^{\prime}$) for \ymgo and \yzgo, respectively. In the insets, $\chi^{\prime}$ in an extended temperature range up to 4~K are plotted. Dashed lines indicate the Curie-Weiss fits for the 100-Hz data. From ref.~\cite{Ma2017Quantum}.
\label{fig4}}
\end{figure}

The spin-glass phase, with frozen, short-range correlations below the freezing temperature $T_{\rm f}$\cite{RevModPhys.58.801,Mydosh1993,Mydosh1986}, can be identified from the a.c. susceptibility. Ma \et have performed such measurements on both \ymgo and \yzgo with temperatures spanning about 3 decades, ranging from 0.05 to 4~K. Some of the results are shown in Fig.~\ref{fig4}. In both compounds, they observe strong frequency-dependent peaks below 0.1~K, evidencing a broad distribution of the spin relaxation times around $T_{\rm f}$, typical for a spin glass\cite{RevModPhys.58.801,Mydosh1993,Mydosh1986,PhysRevB.27.3100,PhysRevB.23.1384,PhysRevB.19.2664,PhysRevB.92.134412}.
They consider disorder and frustration to give rise to the spin-glass phase.

However, a $\mu$SR study on \ymgo\cite{PhysRevLett.117.097201} shows that there is no signature of spin freezing down to 0.07~K, which is already below the $T_{\rm f}$ reported in the a.c. susceptibility measurements\cite{Ma2017Quantum}. One possible origin of this discrepancy is that these two techniques cover different time scales: the former and latter probes are sensitive to fluctuations with frequencies larger and smaller than 10$^4$~Hz, respectively\cite{musrbrochure}. Furthermore, as estimated from the INS results, the portions of the spectral weight in the elastic channel over the total weight, are 16\% and 13\% for \ymgo and \yzgo, respectively\cite{np13_117,Ma2017Quantum}. These roughly represent the portions of moments that have been frozen. All these together may cause difficulties to detect the spin freezing by $\mu$SR.

Another important feature for a spin glass is that in the d.c. susceptibility measurements, there should be a cusp at $T_{\rm f}$, where the zero-field-cooling and field-cooling susceptibility begin to separate from each other. At present, there are no such data available for \ymgo and \yzgo, so performing d.c. susceptibility down to temperatures below 0.1~K will be useful to further clarify the ground state of these compounds.

For organic compounds such as $\kappa$-(BEDT-TTF)$_2$Cu$_2$(CN)$_3$ and EtMe$_3$Sb[Pd(dmit)$_2$]$_2$, disorder effect is expected to be significant, and it is unclear that whether the QSL phase can survive in the presence of strong disorder\cite{doi:10.7566/JPSJ.83.034714,doi:10.7566/JPSJ.83.103704,PhysRevB.92.134407,nature464_199,arcmp2_167,0034-4885-80-1-016502,arcmp5_57,RevModPhys.89.025003,0034-4885-78-5-052502}.
Moreover, as we discuss in Sec.~\ref{kappacu} for $\kappa$-(BEDT-TTF)$_2$Cu$_2$(CN)$_3$, the specific heat indicates a gapless ground state. On the other hand, thermal conductivity measurements reveal no contributions from the magnetic excitations, inconsistent with the gapless QSL state. How to reconcile these contradicting results? Is it because that the disorder effect makes the spinons localized and thus not conduct heat? Or the ground state is not a QSL at all? At the current stage, we believe these are open questions calling for further investigations.

For the most heavily studied kagom\'{e} compound, ZnCu$_3$(OH)$_6$Cl$_2$, disorder also plays an important role. In particular, there are 5-15\% excess Cu$^{2+}$ replacing the nonmagnetic Zn$^{2+}$, which induces randomness in the magnetic exchange coupling\cite{doi:10.1021/ja1070398,PhysRevB.83.100402,PhysRevLett.108.157202}. It is believed that such disorder can be accountable for many experimental observations\cite{doi:10.1021/ja1070398,PhysRevB.83.100402,PhysRevLett.100.087202,PhysRevB.77.052407,PhysRevB.90.064417,PhysRevLett.100.077203,PhysRevLett.100.157205,PhysRevB.76.132411,nm6_853,PhysRevB.78.132406}.
As an example, by considering the Cu impurities, Han \et\cite{PhysRevB.94.060409} estimate a spin gap of 0.7~meV in the kagom\'{e} layer, close to that obtained from the NMR results\cite{science350_655}.

For QSL candidates, frustration is strong. In the presence of strong disorder, the spin-glass phase is often observed, as disorder and frustration are two important ingredients for a spin glass\cite{RevModPhys.58.801,Mydosh1993,Mydosh1986,0295-5075-93-6-67001}. A spin glass mimics a QSL in many aspects---it maintains short-range spin-spin correlations, so in the susceptibility, specific heat, and neutron diffraction measurements, it lacks the signature of a long-range magnetic order; moreover, as demonstrated in ref.~\cite{Ma2017Quantum}, a spin-glass phase can also produce the continuous INS spectra, which is arguably the strongest evidence for a QSL so far. Therefore, in the quest for QSLs, the spin-glass phase which can give rise to spin-liquid-like features must be excluded first before labeling the candidate as a QSL.

Based on discussions above, we now give several perspectives:
\begin{itemize}
\item Although great progress has been made in  theory\cite{nature464_199,arcmp2_167,0034-4885-80-1-016502,arcmp5_57,RevModPhys.89.025003,0034-4885-78-5-052502,0034-4885-74-5-056501,doi:10.1080/14786430601080229,RevModPhys.88.041002,arcmp7_195,RevModPhys.82.53,arcmp3_35,1742-6596-320-1-012004,doi:10.1143/JPSJ.79.011001}, it still lacks a proposal for the defining feature of a QSL that can be detected directly from experiments. At present, observations of the continuous magnetic excitation spectra in INS measurements have often been taken to be the most reliable evidence for a QSL\cite{nature492_406,nature540_559,np13_117}. However, this is a necessary but not sufficient evidence for the fractionalized excitations\cite{Ma2017Quantum}. A feasible direct proposal to identify a QSL should greatly boost this field.

\item As we discuss above, there appear to be no ideal QSLs so far. Materials wise, does there exist a QSL candidate with large magnetic exchange interactions, little disorder, and minimal extra interactions that produce the static magnetic order? In the past, most attention had been paid to materials with triangular or kagom\'{e} lattice where strong geometrical frustration is present\cite{nature464_199}. Now, studying the SOC-assisted Mott insulators with anisotropic bond-dependent Kitaev interactions on the honeycomb lattice may offer new possibilities\cite{Kitaev2006Anyons,prl102_017205,PhysRevB.90.041112}. For instance, very recently, H$_3$LiIr$_2$O$_6$ has been suggested to be a Kitaev QSL\cite{nature554_341}.

\item According to Anderson's proposal, high-temperature superconductivity can emerge from QSLs\cite{PhysRevLett.58.2790,Baskaran1987973,anderson1}. There have been some successes in making QSL candidates superconducting by applying pressures to some organic compounds\cite{PhysRevLett.85.5420,0034-4885-74-5-056501,PhysRevLett.95.177001,PhysRevLett.99.256403,doi:10.1021/ja063525l}.
However, another more common route to achieve superconductivity---via chemical doping, has not been successful so far\cite{PhysRevX.6.041007}. Is it because there is no ideal QSL candidate so far? Will chemical doping an ideal QSL eventually lead to high-temperature superconductivity as
predicted? In this aspect, recent advances in doping using electric-field gating may offer some assistance\cite{doi:10.7566/JPSJ.83.032001,PhysRevLett.116.077002,2011arXiv1104.2119M,nature546_124,CUI201811,HOSONO20185}.
\end{itemize}

\section{Summary}

To summarize, we review recent progress on QSLs, especially on the magnetic-field measurements on several QSL candidates, including the geometrically-frustrated triangular and kagom\'{e} compounds, including \ymgo, \yzgo, $\kappa$-(ET)$_2$Cu$_2$(CN)$_3$, EtMe$_3$Sb[Pd(dmit)$_2$]$_2$, and ZnCu$_3$(OH)$_6$Cl$_2$, and the Kitaev material $\alpha$-RuCl$_3$ with the honeycomb lattice. While there are many experimental evidences showing that they are promising candidates for QSLs, there are also some evidences that may be used to argue against the QSL picture. As such, we provide several perspectives hoping to stimulate further investigations. We anticipate that continuous efforts will be paid off by the discovery of more fascinating physics and ideal candidate materials.

Before ending this review, we note that there are many other materials that have been proposed to be QSLs. We list a few examples below:
\begin{itemize}
\item Na$_4$Ir$_3$O$_8$ is a widely studied QSL candidate with the hyperkagom\'{e} lattice\cite{prl99_137207,PhysRevB.88.220413,PhysRevLett.113.247601,PhysRevLett.118.047201}. In the initial report\cite{prl99_137207} that suggested it to a QSL, spin freezing indicative of a spin-glass phase at $T_{\rm f}=6$~K was observed. The frozen moments were estimated to be less than 10\% of the total moments and were thus ignored. Later on, both $\mu$SR\cite{PhysRevLett.113.247601} and NMR measurements\cite{PhysRevLett.115.047201} showed that the spins are frozen and maintain short-range correlations in the ground state.

\item Kagom\'{e} compounds ZnCu$_3$(OH)$_6$SO$_4$\cite{1367-2630-16-9-093011} and Zn-substituted barlowite
Cu$_3$Zn(OH)$_6$FBr\cite{Feng2017Gapped,Xiao-GangWen:90101}, and a hyperkagom\'{e} material PbCuTe$_2$O$_6$\cite{prb90_035141,prl116_107203}. In Cu$_3$Zn(OH)$_6$FBr, it has been shown that the magnetic field dependence of the
gap extracted from the NMR data is consistent with that given by  fractionalized spin-1/2 spinon excitations\cite{Feng2017Gapped}.

\item A triangular spin-1 material Ba$_3$NiSb$_2$O$_9$\cite{PhysRevB.95.060402}.

\item Ca$_{10}$Cr$_7$O$_{28}$, a system with complex structure, and more interestingly, with ferromagnetic interactions\cite{np12_942,PhysRevB.95.174414,0953-8984-29-22-225802}. In this compound, although Balz $et\,al.$ found that there is no static magnetic order in the $\mu$SR measurements, they observed frequency dependent peaks in the a.c. susceptibility, which is characteristic of a spin glass\cite{np12_942}. However, they argued that the spin-glass phase could be ruled out by doing the Cole-Cole analysis for the a.c. susceptibility data\cite{np12_942}.

\item Very recently, a protypical charge-density-wave compound 1T-TaS$_2$ with the David-star structure has attracted a lot of attention due the possibility of realizing the QSL state\cite{Law03072017,npjqm2_42,np13_1130,PhysRevB.96.195131,PhysRevB.96.081111}.

\end{itemize}
In this review, we do not discuss these materials in details due to the limited space. Readers who are interested in them can refer to the above references and the references therein.

\section{Acknowledgements}
The work was supported by the National
Natural Science Foundation of China with Grant Nos~11674157 and 11822405, and Fundamental Research Funds for the Central Universities with Grant No.~020414380105. We would like to thank our colleagues and collaborators Jian-Xin Li, Shun-Li Yu, Jun-Ming Liu, Shiyan Li, Weiqiang Yu, Xin Lu, P.~\v{C}erm\'{a}k, A.~Schneidewind, Guochu Deng, S. Danilkin, S.~Yano, and J.~S.~Gardner, who made this work possible.



\end{document}